\documentclass[useAMS,usenatbib,usegraphicx]{mn2e}
\usepackage{epsfig}
\usepackage{amsmath} 
\usepackage{rotating}           
\usepackage{color}                
\usepackage{amssymb} 
\usepackage{times}

\def\kms{km ${\rm s}^{-1}$}

\def\ch2{$\chi^2$}

\def\kms {\hbox{${\rm km\ s}^{-1}$}}

\def\scm  {$\hbox{{\rm cm}}^{-2}$}    

\def \AL {$\alpha $}     
\def \HI {H{\sc \,i}}
\def \WpHz {W Hz$^{-1}$}

\def\lapp{\ifmmode\stackrel{<}{_{\sim}}\else$\stackrel{<}{_{\sim}}$\fi}
\def\gapp{\ifmmode\stackrel{>}{_{\sim}}\else$\stackrel{>}{_{\sim}}$\fi}

\title[A survey for \HI\ in the distant Universe]{A survey for \HI\ in the distant Universe: the detection of associated 21-cm absorption at {\boldmath $z=1.28$}}
\author[S. J. Curran et al.]{S. J. Curran$^{1,2}$\thanks{E-mail:
sjc@physics.usyd.edu.au}, M. T. Whiting$^{3}$,   A. Tanna$^{4}$, E. M. Sadler$^{1,2}$,  M. B. Pracy$^{1}$ \newauthor and R. Athreya$^{5}$\\
$^{1}$Sydney Institute for Astronomy, School of Physics, The University of Sydney, NSW 2006, Australia\\
$^{2}$ARC Centre of Excellence for All-sky Astrophysics (CAASTRO)\\
$^{3}$CSIRO Astronomy and Space Science, PO Box 76, Epping NSW 1710, Australia\\
$^{4}$School of Physics, University of New South Wales, Sydney NSW 2052, Australia\\
$^{5}$Indian Institute of Science Education and Research, 900, NCL Innovation Park, Dr Homi Bhabha Road
Pune, Maharashtra 411008, India}

\begin{document}

\date{Accepted ---. Received ---; in original form ---}

\pagerange{\pageref{firstpage}--\pageref{lastpage}} \pubyear{2012}

\maketitle

\label{firstpage}

\begin{abstract}
  We have undertaken a survey for \HI\ 21-cm absorption within the host galaxies of $z\sim1.2 - 1.5$ radio sources, in
  the search of the cool neutral gas currently ``missing'' at $z\gapp1$. This deficit is believed to be due to the
  optical selection of high redshift objects biasing surveys towards sources of sufficient ultra-violet luminosity to
  ionise all of the gas in the surrounding galaxy. In order to avoid this bias, we have selected objects above blue
  magnitudes of $B\sim20$, indicating ultra-violet luminosities below the critical value of $L_{\rm UV}\sim10^{23}$
  \WpHz, above which 21-cm has never been detected.  As a secondary requirement to the radio flux and faint optical
  magnitude, we shortlist targets with radio spectra suggestive of compact sources, in order to maximise the coverage of
  background emission. From this, we obtain one detection out of ten sources searched, which at $z=1.278$ is the third
  highest redshift detection of associated 21-cm absorption to date. Accounting for the spectra compromised by radio
  frequency interference, as well as various other possible pitfalls (reliable optical redshifts and turnover
  frequencies indicative of compact emission), we estimate a detection rate of $\approx30$\%, close to that expected for
  $L_{\rm UV}\lapp10^{23}$ \WpHz\ sources.

\end{abstract}


\section{Introduction}
\label{intro}
 
\subsection{A paucity of cool neutral gas at high redshift}

Despite the detection of large columns of neutral hydrogen in the early Universe, through the known 1500 damped
Lyman-$\alpha$ absorbers (DLAs, see \citealt{cwbc01,npls09} and references therein)\footnote{Where the neutral hydrogen
  column density exceeds $N_{\rm HI}=2\times10^{20}$ \scm.}, at only 80, detections of the 21-cm transition of neutral
atomic hydrogen at $z\geq0.1$ remain rare.\footnote{The low probability of the transition, compounded by the inverse
  square law, renders 21-cm in undetectable in emission at $z\gapp0.1$ (see \citealt{chg+08}).  In absorption however,
  the strength of the line only depends upon the column density and the flux of the continuum
  source, allowing this to be detected to much higher redshift.} Detection of this at high redshift is of particular importance since this transition traces the cool component
of the neutral gas, the reservoir for star formation, whereas the Lyman-\AL\ transition traces all of the neutral gas.

The paucity of redshifted 21-cm absorption to date can largely be attributed
to the low fraction of background sources with sufficient radio flux ($\gapp0.1$ Jy), the narrow bandwidths free of
radio frequency interference (RFI), as well as the limited frequency coverage of current radio
telescopes.
Of the detections, half are due to absorbers {\em intervening} the sight-lines to more distant
radio sources (see \citealt{gsp+12} and references therein), with the other half  being {\em associated} with the 
gas in the host galaxy of the source itself  (see \citealt{cw10} and references therein). Furthermore, both types of
absorber are subject to additional selection effects, increasing the difficulty of detection at high redshift:
\begin{itemize}
\item[--] For the intervening absorbers, the majority of detections occur at redshifts of $z\lapp2$ (see
  \citealt{cur09a}) and for those known to contain large columns of neutral hydrogen (the DLAs), the detection rate at
  $z\gapp2$ is only a third that at $z\lapp2$ (60\%, cf. 20\%). This is believed to be due to the geometry effects
  introduced by an expanding Universe, where high redshift absorbers are always disadvantaged in how effectively they
  intercept the background emission \citep{cur12}.
\item[--] For the associated absorbers, at lower redshifts ($z\lapp1$) there is a $\approx40\%$ probability of a
  detection, depending on whether the large-scale gas disk intercepts the line-of-sight to the radio source
  \citep{cw10}. However, at higher redshifts the probability of a detection falls drastically. This is believed to be
  due to the traditional optical selection of targets, where the availability of a redshift biases towards the most ultra-violet luminous
  sources \citep{cww+08}. For most of the radio sources known 
at these large luminosity distances all of the neutral gas is probably ionised \citep{cw12}.
\end{itemize}
This highlights a major difference between the two types of absorber: While there are equally few detections of 21-cm in
intervening systems, all of these are known or suspected\footnote{Having a rest-wavelength of $\lambda=1216$ \AA, the
  Lyman-\AL\ transition can only be detected by ground-based telescopes at redshifts of $z\gapp1.7$. At $z\lapp1.7$, the
  Mg{\sc \,ii} $\lambda=2796$ \AA\ equivalent width may be used infer the presence of a DLA \citep{rtn05} and perhaps
  even to estimate the total hydrogen column density \citep{cur09a}.}  to contain large columns of neutral hydrogen and,
where searched, 21-cm detection rates are high \citep{gsp+12}.  However, within the high redshift radio sources
themselves, the neutral gas may be truly absent (the ``UV interpretation'', \citealt{cw12}) and until a population of
21-cm absorbing galaxies is found, this material, the reservoir for star formation, may be regarded as ``missing''.
Finding these missing systems is of utmost importance in quantifying the number of gas-rich galaxies which exist below
the detection thresholds of optical spectroscopy. Based upon the UV interpretation, we therefore suggest that high
redshift surveys should be directed towards the most optically dim radio sources, where the faint magnitudes indicate
that the ultra-violet luminosity of the active galactic nucleus (AGN) is below the critical ionising
continuum luminosity of $L_{\rm UV}\sim10^{23}$ \WpHz, above which associated 21-cm absorption has never been detected
(\citealt{cww+08,cw10, cwm+10,gd11}).\footnote{\citet{psv+12} also find a critical X-ray luminosity, above which sources
  are not detected in 250 $\mu$m continuum
  emission.} 

\subsection{Source selection}
\label{ss}

\begin{figure*}
\centering \includegraphics[angle=-90,scale=0.83]{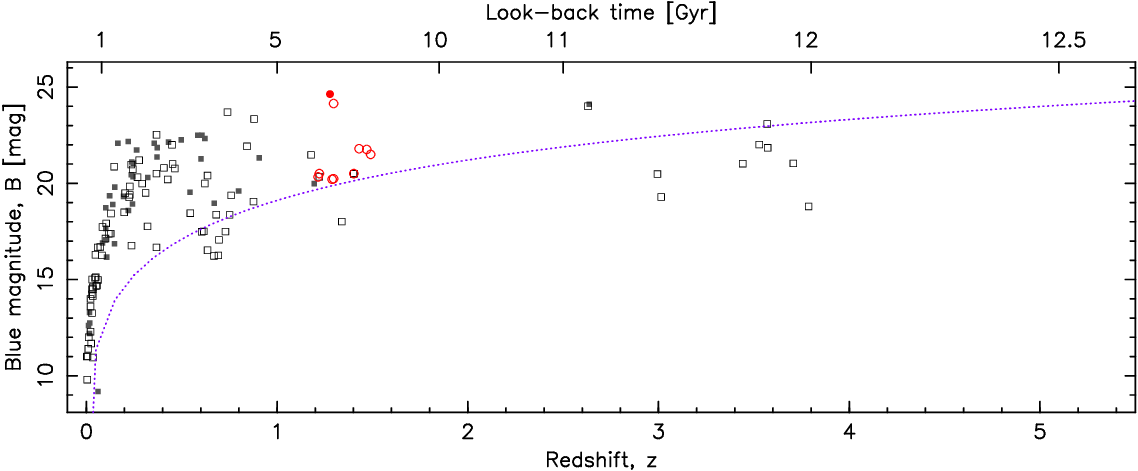}
\caption{The blue magnitudes of the redshifted associated 21-cm searches (compiled in \citealt{cw10,ace+12}). The filled
  symbols show the detections and the unfilled symbols the non-detections, with the squares showing the previous targets and the
  circles our targets (all but one are non-detections, Sect. \ref{observations}).  The curve shows which $B$ magnitude
  corresponds to the critical $L_{\rm UV}=10^{23}$ \WpHz, above which all of the gas will be ionised,  for a spectral slope of 
$\alpha = -1.5$, used to select the targets (cf. figure 5 of \citealt{cww09} and figure 1 of \citealt{cwsb12}). 
The look-back time is calculated for $H_{0}=71$~km~s$^{-1}$~Mpc$^{-1}$, $\Omega_{\rm matter}=0.27$ and
$\Omega_{\Lambda}=0.73$ (used throughout the paper).}
\label{B-z}
\end{figure*}

Ideally, we would like to search for associated 21-cm absorption at redshifts of $z\gapp3$, the regime of the original exclusive non-detections which
alerted us to the critical luminosity \citep{cww+08} and where there is currently only one detection  (at $z=3.4$, \citealt{ubc91}).  
However, our usual source catalogues, the Parkes Half-Jansky Flat-spectrum Sample (PHFS, \citealt{dwf+97}) and
Quarter-Jansky Flat-spectrum Sample (PQFS, \citealt{jws+02}), yield only two sources with suitably faint blue magnitudes
($B\gapp22$, cf. Fig. \ref{B-z}) in the 90-cm band ($z=3.09 - 3.63$), both of which have been previously searched (\citealt{cww+08}).
Furthermore, the remaining catalogues of radio sources, which have both magnitudes and redshifts available
\citep{cwwa11}, yield only ultra-steep spectrum sources in the 305--360 MHz band ($z = 2.9 -3.7$). The steep radio
spectra may indicate very extended emission, minimising the covering factor, $f$, of the neutral gas, possibly a cause
of the exclusive non-detections of $L_{\rm 1216}\lapp10^{23}$ \WpHz\ sources at these redshifts \citep{cwsb12}.

In order to bypass these issues, we are forced to probe lower redshifts ($z\sim1.2 - 1.5$), where 21-cm is redshifted into the 610 MHz
band of the Giant Metrewave Radio Telescope (GMRT).\footnote{http://gmrt.ncra.tifr.res.in/} In order to avoid steep spectrum sources, we have therefore
prioritised the faintest objects mostly from the PHFS and PQFS, where the blue magnitudes of $B\gapp20$ indicate luminosities of $L_{\rm
  1216}\lapp10^{23}$ \WpHz\ at these redshifts (Fig.~\ref{B-z}). These catalogues, by definition, comprise of sources with flat
spectra over 2.7--5.0 GHz, although in order to obtain a sample of ten, a few were added from the other radio catalogues
(see Table~\ref{obs}), with the priority of $B\gapp20$ and a flux density in excess of 0.1~Jy, meaning that a target with a flat spectrum or a 
clear turnover frequency  
could not always be included. From a survey of these ten objects, we
report the third highest redshift detection of associated 21-cm absorption to date (at $z=1.278$ in SDSS\,J154508.52+475154.6), 
as well as eight upper limits and one source lost to RFI, which we present and discuss here.

\section{Observations}
\label{observations}

Each of the sources was observed on 22--24 January 2012 with the GMRT full 30 antenna array, using the
610 MHz receiver backed with the FX correlator over a bandwidth of 16 MHz. 
This was spread over 512 channels in orthogonal circular polarisations, giving a channel spacing of $\approx15$
\kms. This is sufficient to spectrally resolve all of the associated 21-cm absorbers currently known (which range from 18 to 475 \kms, \citealt{cwsb12}),
while maintaining a redshift coverage of $\Delta z \approx \pm0.03$, in order to cover any uncertainties in the optical redshifts
of the targets.  Each source was observed for a total
of two hours with the aim of reaching a $3\sigma$ optical depth limit of $\tau\lapp0.01$ per
channel, or a sensitivity to $N_{\rm HI}\lapp2\times10^{17}.\,(T_{\rm spin}/f)$ \scm, which is close to the lower limit
for the published 21-cm searches. 

\begin{figure}
\vspace{6.8cm} 
	\setlength{\unitlength}{1in} 
\begin{picture}(0,0)
\put(-0.5,-1.5){\includegraphics{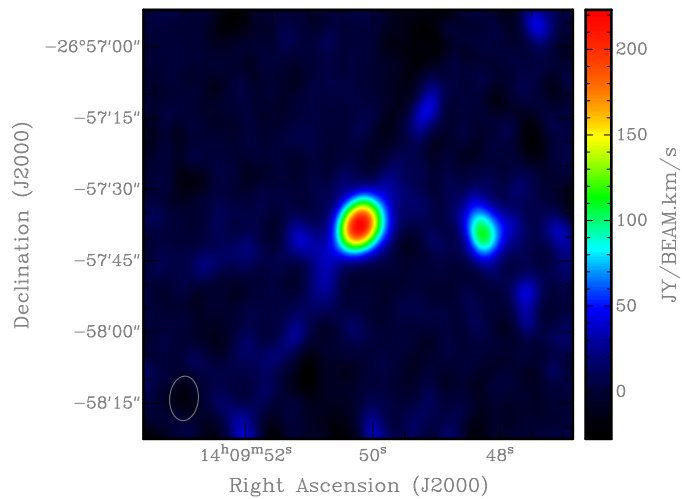}} 
\end{picture}\caption{GMRT 585 MHz continuum image of J1409--2657. The $9.4"\times6.0"$ synthesised 
  beam is shown in the bottom left corner.}
\label{map}
\end{figure}

For each source, 3C\,48, 3C\,147 and 3C\,286 were used for bandpass calibration and a strong nearby point source for
phase calibration.  The data were flagged and reduced using the {\sc miriad} interferometry reduction package, with
flagging of the edge channels leaving the central 470 channels, giving a span of $\approx\pm4000$ \kms.  After averaging
the two polarisations, a spectrum was extracted from the cube. As per usual when reducing GMRT data with {\sc miriad},
it was not possible to flux calibrate the targets via the {\sf gpboot} task. The flux densities of the (mostly
unresolved) sources were found to be generally lower than those estimated (Table \ref{obs}),
although the derived  optical depths will be accurate. Regarding each source:\\
{\bf PKS\,0220--349} was observed for a total of 2.58 hours. After flagging of non-functioning antennas (22 and 23), 
the RR polarisation seemed worst affected be RFI. Further flagging left 145 baseline pairs in this polarisation 
and 378 in the LL.  The source was partially resolved by the $10.9"\times6.8"$ synthesised beam.\\
{\bf BZQ\,J0301+0118} was observed for a total of 2.08 hours. 
All antennas were functioning, giving a high quality image from the 435 baselines pairs.  Upon the extraction
of a spectrum, however,  the bandpass was dominated by spikes and flagging of the worst affected
data, leaving 374 baseline pairs, revealed a sinusoidal ripple. Further flagging reduced this somewhat, although it
was still evident. The source was partially resolved by the $6.8"\times5.0"$
synthesised beam.\\ 
{\bf PKS\,0357--264} was observed for a total of 1.53 hours. 
All antenna were functioning, but severe RFI meant that no reasonable image nor spectrum could be produced. Flagging of
the baselines in which r.m.s. noise level exceeded 1 Jy, left only 12 baseline pairs from which the image and spectrum
fared no better.\\ 
{\bf PKS\,0400--319} was observed for a total of 2.01 hours. 
RFI meant that all but the first 30 minutes of the data had to be flagged on all baseline pairs,
before a satisfactory image could be produced, although the extracted spectrum still exhibited spikes. The source
was unresolved by the $9.1"\times5.8"$ synthesised beam.\\
{\bf PKS\,0511--220} was observed for a total of 2.07 hours. All antennas were functioning, giving
a high quality image from the 435 baselines pairs. The extracted spectrum
was, however, dominated by spikes, due to RFI throughout the observing run. We therefore flagged all baselines pairs for
which the r.m.s. noise level exceeded 1 Jy, leaving 147 pairs. The extracted spectrum, although noisier, was far less
dominated by the spikes which marred the unflagged spectrum.  The source was unresolved by the $10.7"\times5.4"$
synthesised beam. \\ 
{\bf [HB89]\,1004--018} was observed for a total of 2.38 hours. All antennas were functioning, although 
baselines consisting of antenna 7 and a
number of other baselines pairs, which exhibited a ripple, were flagged. This left 401 baseline pairs
and the source was unresolved by the $6.3"\times5.3"$ synthesised beam. \\ 
{\bf CGRaBS\,J1334--1150} was observed for a total of 1.89 hours. All antenna were functioning and only the LL
polarisation of antenna 17 had to be removed. Following this, a satisfactory image was produced from which the extracted
spectrum showed evidence of absorption with a peak of $z = 1.402292$, i.e. redshifted by 40 \kms\ and within $\Delta z =
0.0001$ of the optical redshift ($z = 1.402$). However, since this, and other similar features of a single channel
width, appear at various locations in the image, we believe that this is a birdie. The source was unresolved by the
$7.4"\times6.0"$ synthesised beam. \\ 
{\bf CGRaBS\,J1409--2657} was observed for a total of 2.04 hours. 
Although all antennas were functioning, phase calibration was not possible until the LL polarisation was removed. Upon
this, an image could be produced and further flagging of bad baselines (leaving 349 pairs) revealed a double source
(Fig. \ref{map}). Although some other artifacts remain in the image, further {\sc clean}ing of the data was
detrimental. \\
{\bf 4C\,+04.51 (PKS\,1518+047)} was observed for 1.66 hours, although RFI meant that the last 40 minutes has to be removed. All antenna
were functioning and, due to the strength of the source,  self calibration produced a high quality image, although some spikes are still
evident in the bandpass. The source was unresolved by the $6.3"\times5.1"$ synthesised beam. \\ 
{\bf SDSS\,J154508.52+475154.6} was observed for 2.07 hours. Flagging of non-functioning antennas (22, 23 and 30), left
351 baseline pairs, from which an absorption line was clearly apparent in each polarisation of the averaged spectra.
The source was unresolved by the $9.9"\times6.4"$ synthesised beam. \\

\begin{figure*}
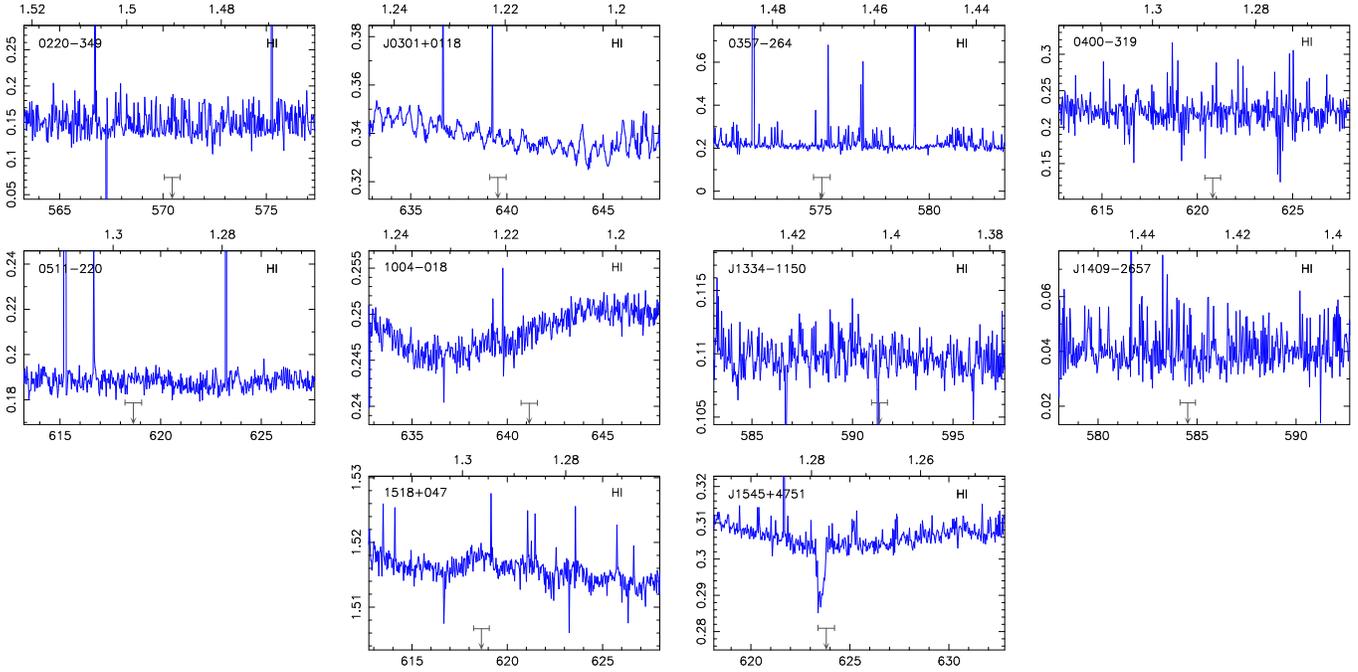
  
\vspace{9.2cm}  
\includegraphics{PKS0220-349.LTA-phase_378+145.dat.ps}  
\includegraphics{bzqj0301.640_phase_374.dat.ps} 
\includegraphics{0357-264_435.dat.ps} %
\includegraphics{0400-319_first0.5hr-phase.dat.ps}
\includegraphics{0511-220_147.dat.ps} 
\includegraphics{1004-018_phase_406.dat.ps}  
\includegraphics{1334-1150_phase_434.dat.ps} 
\includegraphics{J1409-2657_phase_RR_349.dat.ps} 
%
\includegraphics{4C+04.51-self.dat.ps} 
\includegraphics{J1545+4751_phase.dat.ps} 
 \caption{The full-band extracted spectra. The ordinate gives the flux density [Jy] and the abscissa the barycentric
  frequency [MHz]. The downwards arrow shows the expected
frequency of the absorption from the optical redshift, with the horizontal bar showing a span of $\pm200$ \kms\ for guidance 
(the mean profile width of the 21-cm detections is 167 \kms). 
The scale along the top axis shows the redshift of \HI\ 21-cm. }
\label{spectra}
\end{figure*}

In Fig. \ref{spectra} we show the final spectra and summarise the results in Table \ref{obs}. 
\begin{table*} 
\centering
\begin{minipage}{175mm}
  \caption{The strong radio sources with $B\gapp20$, in which \HI\ 21-cm falls into the 570--650 MHz band, searched with
    the GMRT.  $z$ is the optical redshift of the source, $S_{\rm est}$ is the flux density estimated by interpolating between
    0.4 and 1.4 GHz (except for J1406--2657 which is extrapolated between 1.4 and 2.7 GHz), $S_{\rm meas}$ the measured
    flux density [Jy], $\Delta S$ is the peak depth of the line/r.m.s. noise [Jy] reached per $\Delta v$ channel [\kms] and $\tau$ the
    derived optical depth, where $\tau=-\ln(1-3\Delta S/S_{\rm cont})$ is quoted for the non-detections. $N_{\rm HI}$ is
    the resulting column density [\scm~K$^{-1}$], where $T_{\rm spin}$ is the spin temperature and $f$ the 
    covering factor, followed by the redshift range over which the limit applies.  Finally we list the $B$ magnitude,
    which suggested $L_{\rm UV}\lapp10^{23}$ \WpHz\ for each source, followed by the catalogue reference.}
\begin{tabular}{@{}l l c   c c r  r c c c  c  @{}} 
\hline
\smallskip
Source                  &  $z$  &$S_{\rm est}$ &  $S_{\rm meas}$  &$\Delta S$ & $\Delta v$ & $\tau$ & $N_{\rm HI}.\,(f/T_{\rm spin})$ & $z$-range & B&  Ref \\
\hline
PKS\,0220--349 &  1.49 & 0.58 &   0.146 &  $0.019$ &  17.1&  $<0.48$ &  $<2\times10^{19}$ & 1.470--1.503   &   21.50 &   D97 \\
BZQ\,J0301+0118  & 1.221 & 0.74&   0.340 &  $0.007$ & 15.3 & $<0.06$ & $<2\times10^{18}$ & 1.192--1.245  &        20.5 &   J02 \\
PKS\,0357--264 & 1.47 & 1.26 &   --- & \multicolumn{4}{c}{\sc no limits available due to rfi}  &  1.434--1.493&21.76 &    D97 \\
PKS\,0400--319 & 1.288  & 0.65 & 0.219&  $0.019$&   15.7  &  $<0.30$    & $<9\times10^{18}$& 1.262--1.318 &20.21 & D97 \\
PKS\,0511--220 & 1.296  & 0.61 & 0.187 & $0.004$ &   15.8  &  $<0.07$    & $<2\times10^{18}$ &1.279--1.302&20.24 & D97 \\
\protect[HB89]\,1004--018   & 1.2154 & 0.60 & 0.248&  $0.003$&   15.2  &  $<0.04$    & $<1\times10^{18}$ &1.192--1.244&20.33& D97 \\
CGRaBS\,J1334--1150   & 1.402  &  0.37 & 0.109&  $0.002$$^{a}$&    16.5&   $<0.06$   & $<2\times10^{18}$ &1.376--1.436&20.5 &   J02 \\
CGRaBS\,J1409--2657  & 1.43$^{b}$ &  0.10 & 0.039&  $0.004$&  16.7   &   $<0.37$   & $<1\times10^{19}$&1.396--1.454&21.8$^{c}$ & W92,D97,J02 \\
4C\,+04.51  & 1.296  &   4.94 & 1.515& $0.002$&    15.8  &  $<0.004$  &  $<1\times10^{17}$$^{d}$ & 1.263--1.317&24.14$^{e}$  & W85,S96\\
SDSS\,J154508.52+475154.6   & 1.277   & 0.76 & 0.304& $0.019$$^{f}$ & 15.6&  0.065   &   $1.8\times10^{19}$   & 1.2780&24.63$^{g}$  & T94,M98 \\
\hline       
\end{tabular}
\label{obs}  
{References: W85 -- \citet{wp85}, W92 -- \citet{whi92a}, T94 -- \citet{tvp+94}, S96 -- \citet{sk96b}, D97 -- \citet{dwf+97} [PHFS],  M98 -- \citet{mac98}, J02 -- \citet{jws+02} [PQFS].\\
Notes: $^{a}$Noise level is valid over region excluding the absorption spike, redshifted by 40 \kms\ ($z =1.4023$) with respect to the reference redshift. $^{b}$Quoted value of $z=2.43$ \citep{dwf+97} is
probably incorrect (Sect. \ref{of}). $^{c}$$B$ magnitude from J02 (D97 quote $B=20.30$). $^{d}$21-cm also undetected by \citet{gss+06} to a weaker limit [$N_{\rm HI} <6\times10^{17\,}(T_{\rm spin}/f)$ \scm]. $^{e}$Estimated from $V=22.80$ \citep{wp85}, using the fits in \citet{cw10}.  $^{f}$The peak depth of the absorption. The r.m.s. noise is  0.002 Jy. $^{g}$Estimated from $R=21.76$ \citep{mac98}, using the fits in \citet{cw10}. 
}
\end{minipage}
\end{table*} 

\section{Results and discussion}
\subsection{Detection of associated \HI\ 21-cm at {\boldmath $z = 1.278$}}

\begin{figure}
\centering 
\includegraphics[angle=270,scale=0.45]{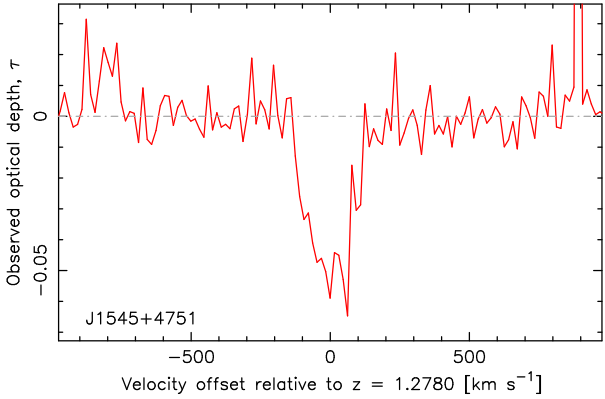} 
\caption{\HI\ 21-cm  absorption in SDSS\,J154508.52+475154.6. The ordinate shows the observed optical depth, 
for a continuum level of 0.304~Jy, and the 
abscissa velocity offset from the mean-weighted redshift of  $z_{\rm mean} = 1.2780$. }
\label{J1545+4751}
\end{figure}

The survey has resulted in a clear detection of redshifted \HI\ 21-cm absorption, which we show in Fig. \ref{J1545+4751}.
The deepest feature has a redshift of $z_{\rm peak} = 1.27851$, which is redshifted by 65 \kms\ from the mean-weighted value
of $z_{\rm mean} = 1.27802\pm0.00002$, which itself is redshifted by 134 \kms\ from the reference optical value of $z_{\rm opt}=1.2772\pm0.002$ \citep{vt95}.
The feature spans 280 \kms, with a full-width half maximum of FWHM$\,=151\pm8$ \kms,  giving an 
integrated optical depth of  $\int\tau dv = 9.69\pm 0.53$ \kms. This yields a column density of 
$N_{\rm HI} = 1.77\pm0.10\times 10^{19}\,(T_{\rm spin}/f)$ \scm, which for a modest $T_{\rm spin}/f\gapp10$~K 
(cf. the lowest $T_{\rm spin}/f= 60$~K yet found, \citealt{ctm+07}), qualifies
this as a damped Lyman-$\alpha$ absorber. 
The highest column density of both the SDSS DR5 and DR7 DLAs is $N_{\rm HI} = 8\times 10^{21}$ \scm\ (\citealt{phh07,npls09}, respectively), which the new detection attains
for $T_{\rm spin}/f = 450$~K, well below the mean value of $T_{\rm spin}/f= 1800$~K \citep{cur12}.

\subsection{Properties of the sample affecting the detection of 21-cm}
\subsubsection{Sensitivity limits}
\label{sl}

\begin{figure*}
\centering 
\includegraphics[angle=270,scale=0.65]{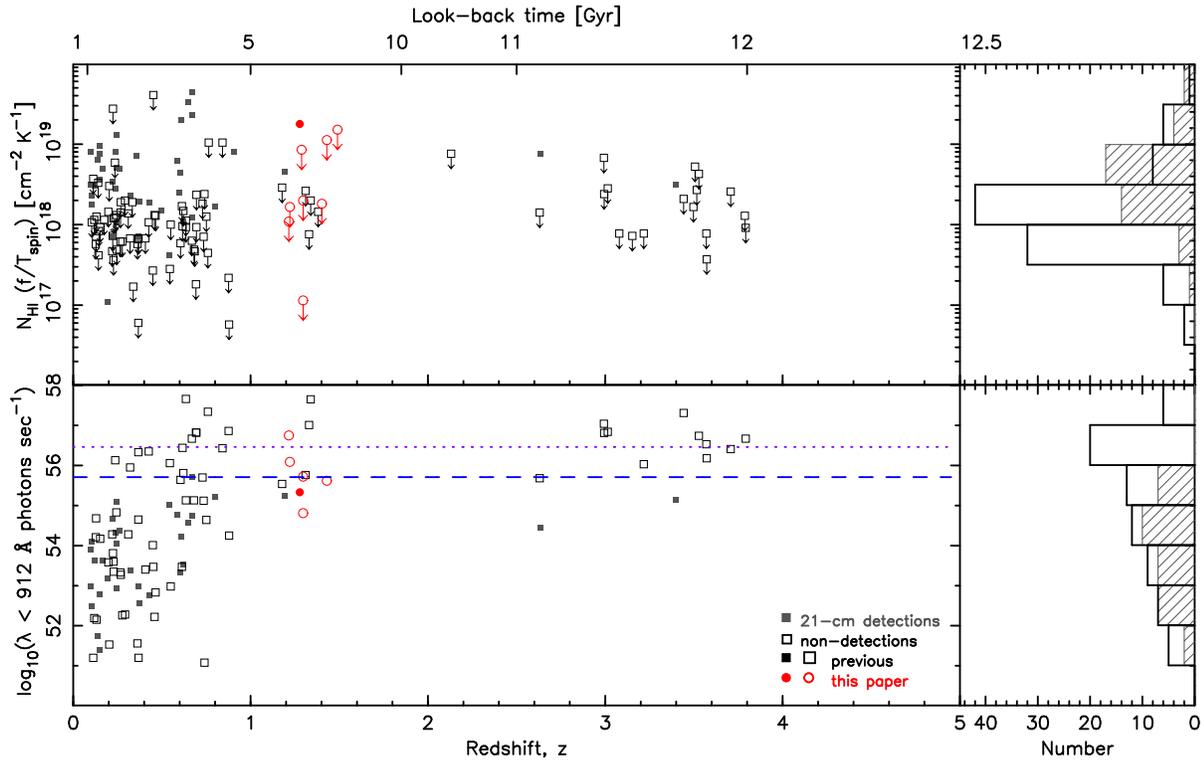}
\caption{The normalised ($1.823\times10^{18}\int\tau\,dv$) 21-cm line-strength (top) and the 
ionising photon rate  (bottom)  versus redshift for the $z\geq0.1$ quasars and
radio galaxies searched in associated 21-cm absorption.  
The filled symbols/hatched histogram represent the 21-cm detections and the unfilled symbols/unfilled histogram the non-detections
(where these are $3\sigma$ upper limits in the top panel). The horizontal lines in the bottom panel show
the highest photon rate at which there is a 21-cm detection ($\int^{\infty}_{\nu_{\rm ion}}({L_{\nu}}/{h\nu})\,d{\nu} = 5.1\times10^{55}$ s$^{-1}$ -- broken) and 
the critical value estimated from the composite SEDs ($2.9\times10^{56}$ s$^{-1}$ -- dotted, see \citealt{cw12}).
The previous sources 
are shown as squares (listed in \citealt{cw10}, with the addition of those in \citealt{cwm+10}), with the results presented here
shown as circles. }
\label{lum-z}
\end{figure*}

In Fig. \ref{lum-z} (top panel) we show the line strength of the detection and the limits for the eight non-detections,
in context of the previous searches. From this we see that approximately half of our non-detections 
have been searched to reasonable limits, i.e. $N_{\rm
  HI}.\,(f/T_{\rm spin})\lapp 10^{18}$ \scm~K$^{-1}$ and so these are sensitive to the majority of the 21-cm detections.
One out of nine targets is a detection rate of only $\approx10$\%, although considering that three 
(0220--349, 0400--319 \& J1409--2657)   have $3\sigma$ limits to only $N_{\rm HI} \gapp10^{19}\,(T_{\rm spin}/f)$ \scm, one out
of six (a rate of $\approx17$\%) could be more appropriate. This is, however, still low in comparison to the $\approx40$\%  generally found 
at $L_{\rm UV}\lapp10^{23}$ \WpHz\ \citep{cw10}, this rate arising from the likelihood of the large-scale gas disk being orientated so that it
 intercepts our line-of-sight to the AGN. We discuss the possible reasons for this low detection rate here.

\subsubsection{Ionisation of the neutral gas}
\label{ionng}
\begin{figure*}
\centering \includegraphics[angle=-90,scale=0.62]{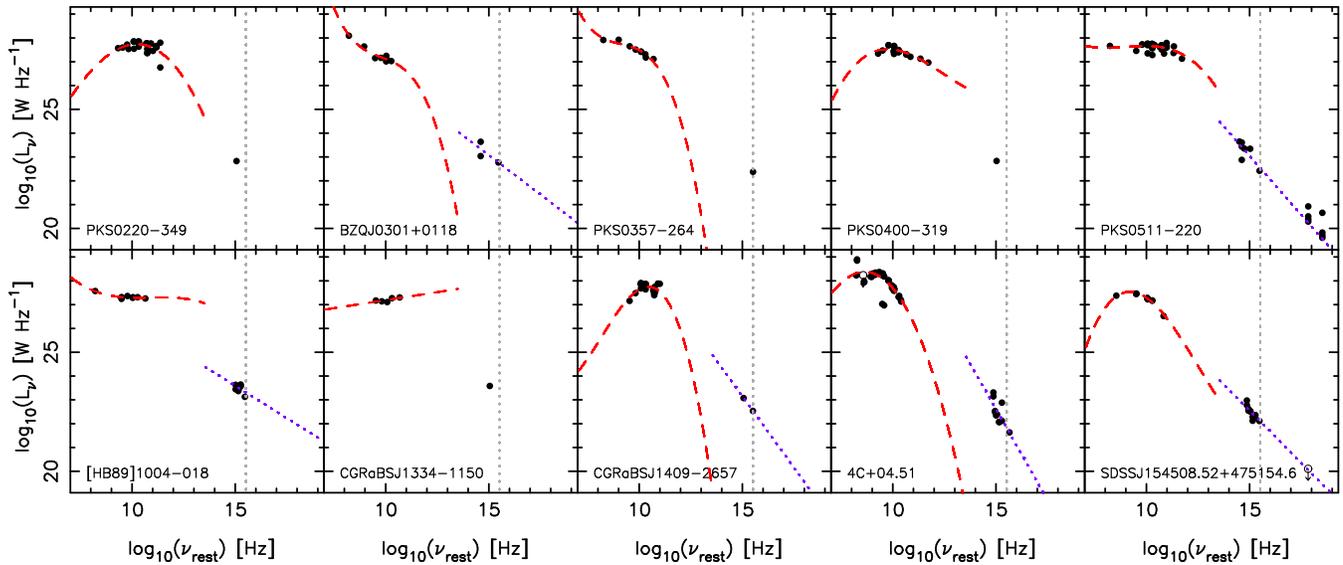}
\caption{The rest-frame SEDs of our targets overlaid by fits to the photometry.
The broken curve shows the third order polynomial fit to the radio data ($\nu_{\rm rest} \leq10^{12}$ Hz) and the dotted line the power-law fit to the UV data ($10^{14.5}\leq \nu_{\rm rest} \leq10^{17}$ Hz), 
where possible (see \citealt{cwsb12} for details). For the sake of clarity both have been extended to $10^{13.5}$ Hz.
The vertical dotted line signifies a rest-frame frequency of $3.29\times10^{15}$ Hz ($\lambda = 912$~\AA).}
\label{lum-freq}
\end{figure*}

Although there is an apparent critical luminosity of $L_{\rm 1216}\sim L_{\rm 912}\equiv L_{\rm UV}\sim10^{23}$ \WpHz\ above which 21-cm has never
been detected \citep{cww+08,cw12}, the gas is ionised by all photons of wavelength $\lambda\leq912$ \AA. We are therefore 
interested in  the total ionising luminosity, $\int^{\infty}_{\nu_{_{\rm ion}}}(L_{\nu}/\nu)\,d{\nu}$, where $\nu_{_{\rm ion}} = 3.29\times10^{15}$ Hz,
 which we derive from the SED fits to the photometry (Fig. \ref{lum-freq}).\footnote{See \citet{cwsb12} for a detailed description 
of the polynomial fits to the SEDs.}
Dividing the total ionising luminosity by the Planck constant, i.e. $\int^{\infty}_{\nu_{_{\rm ion}}}(L_{\nu}/h\nu)\,d{\nu}$, gives
the ionising photon rate \citep{ost89}, which we show in
Fig. \ref{lum-z} (bottom panel) and summarise in Table \ref{lums}. From this, we see that the 21-cm detection (J1545+4751)
arises in the second least luminous source (the faintest being 4C\,+04.51, which also has the strongest limit, Table~\ref{obs}).

Note that these are two of only three sources below $5.1\times10^{55}$ s$^{-1}$, the highest photon rate at which
there has been a 21-cm detection (in 3C\,216, \citealt{vpt+03}), the third being J1409--2657 which has been searched to a poor limit [allowing 
$N_{\rm HI} \leq 1.1\times10^{19}\,(T_{\rm spin}/f)$ \scm].  Of the remaining six targets, three are above 
$5.1\times10^{55}$ s$^{-1}$
 and photon rates cannot
be determined for the other three (Fig. \ref{lum-freq}), 
although all but one of the six for which photon rates could be
determined are below the critical value of $2.9\times10^{56}$ s$^{-1}$ (estimated from the critical observed luminosity of $L_{\rm 912} = 1.1\times10^{23}$ \WpHz\ and the
composite UV spectral index of the sources above this luminosity, \citealt{cw12}). 
If we exclude J1409--2657 on the basis of its poor limit, applying $5.1\times10^{55}$ s$^{-1}$ as the cut-off
gives a $50$\% detection rate and $2.9\times10^{56}$ s$^{-1}$ a 25\% rate, below the cut-off. Both
are close to the $\approx40\%$ expected for $L_{\rm UV}\lapp10^{23}$ \WpHz\ sources.  
Given that a photon rate could not be determined for J1334--1150, although this has been searched to $N_{\rm HI} <2\times10^{18}\,(T_{\rm spin}/f)$ \scm,
our detection rate could be lower (see Sect. \ref{of} for further discussion).

\subsubsection{Coverage of the background flux as traced through radio properties}
\label{cov}

From the above, it is possible that the detection rate is as high as 50\%, if the highest photon rate at which 21-cm
has been detected is close to the critical value at $z\gapp1$: The value of $3\times10^{56}$ s$^{-1}$ is sufficient
to ionise all of the neutral gas in a large spiral \citep{cw12} and so 
any evolution in 
galactic morphology with redshift, where there is a larger fraction of smaller galaxies at large look-back times,
could lower this critical value \citep{cwsb12}.  That is,  since our targets are at look-back times which are 
at approximately half the age of the Universe (Fig. \ref{lum-z}), a lower critical value may be applicable.

Nevertheless, \citet{cwsb12} cannot rule out steep radio spectra, indicating extended sources, as a cause
of their exclusive non-detections at $z\gapp2$. This motivated the preference for SEDs which 
exhibit a clear turnover for this survey. To verify that our targets fulfil this requirement, as described in 
\citet{cwsb12}, we fit a third order polynomial to the $\lapp10^{12}$ Hz photometry in order to obtain
a radio SED (Fig. \ref{lum-freq}), from which we obtain the turnover frequency and a spectral index (Table \ref{lums}).
\begin{table} 
\centering
\begin{minipage}{85mm}
  \caption{Radio properties and luminosities of the targets. $\nu_{_{\rm TO}}$ is the
  rest-frame turnover frequency, where the upper limits designate no observed turnover (which is thus assumed to
  occur below the observed frequencies), followed by the spectral index at the rest-frame 21-cm, \AL. In the last
  three  columns, we list the $\lambda=1216$ \AA\ rest-frame continuum luminosity and 
  the ionising ($\lambda\leq912$ \AA) photon rate, both corrected for Galactic extinction (see \citealt{cw12}).}
\begin{tabular}{@{}l r r  c  c   @{}} 
\hline
\smallskip
Source                & $\nu_{_{\rm TO}}$ [GHz]  & $\alpha$ & $L_{1216}$ [\WpHz] &  rate  [s$^{-1}$]\\
\hline
0220--349 &   17  &  0.51&  --- & --- \\
J0301+0118  &  $<0.17$ &  $-0.46$  & $6.6\times10^{22}$  &   $1.2\times10^{56}$ \\
0357--264 &    $<1.0$ &  $-0.21$  &--- & --- \\
0400--319 &  5.9 & 0.30   & ---   & --- \\
0511--220 &   8.5& 0.06  &$4.5\times10^{22}$ &   $5.2\times10^{55}$\\
1004--018   &  $<0.16$  &  $-0.13$ & $2.4\times10^{23}$ &   $5.6\times10^{56}$\\
J1334--1150   &  ---   & $0.13$  &    --- & --- \\
J1409--2657  &  28  & 1.13& $4.7\times10^{22}$ &  $4.2\times10^{55}$\\
4C\,+04.51  &   0.45   & $-0.37$   &  $1.0\times10^{22}$ &   $6.5\times10^{54}$ \\
J1545+4751  &   0.89 & $-0.09$  &  $1.6\times10^{22}$  &   $2.1\times10^{55}$ \\
\hline       
\end{tabular}
\label{lums}  
\end{minipage}
\end{table} 

From this, we see that six of the targets have clear turnover frequencies in the observed bands, of which the detection
is one, this also having the second flattest spectral index. For the previous $z\gapp2$ survey \citep{cwsb12}, the mean
turnover frequency was $\left<\nu_{_{\rm TO}}\right> \approx0.08$ GHz (cf. $\approx0.15$ GHz for the 21-cm detections)
and the spectral index $\left<\alpha \right>\approx1.0$ (cf. $\approx0.3$ for the detections). As required, our targets
generally have higher turnovers ($\left<\nu_{_{\rm TO}}\right> \approx3.1$ GHz for all and 1.6 GHz for those with limits
of $N_{\rm HI} \lapp10^{19}\,(T_{\rm spin}/f)$ \scm) and flatter spectra ($\left<\alpha \right>\approx0$ for all and
$-0.2$ for those with limits of $N_{\rm HI} \lapp10^{19}\,(T_{\rm spin}/f)$ \scm). This indicates that these are 
more compact sources \citep{ffs+90} and so should generally cover the background emission at least as effectively as those
previously detected in 21-cm.

Lastly, it is interesting to note that, from VLBA observations,  the detected source is known to be very compact -- 13.3 mas at 5 GHz
\citep{htt+07}. This corresponds to a linear size of just 112 pc at $z=1.278$. 5 GHz is 11 GHz in the rest-frame and assuming that
the source is an order of magnitude larger at 1 GHz, means that an absorber of diameter $\sim1$ kpc would be sufficient to 
intercept all of the emission. Unfortunately, high resolution radio maps are not available for the other targets, due in part to the
selection of a large proportion of southern sources.

\subsubsection{Excitation of the gas by radio emission}

As well as being below the critical photon rate and having radio SEDs indicative of a compact radio source, we note that
the detection has the lowest radio continuum luminosity 
 of the sample ($L_{\rm radio} = 3\times10^{27}$ \WpHz\ at 1.4 GHz, Fig. \ref{lum-freq}). As well as ionisation of the gas,
excitation to the upper hyperfine level by $\lambda\leq 21$ cm photons, raising the spin temperature of the gas \citep{pf56}, may
also cause the gas to be undetectable in 21-cm absorption. Unlike the $\lambda=1216$~\AA\ continuum luminosity, \citet{cww+08} found 
no correlation between the $\lambda = 21$ cm continuum luminosity and the incidence of detection, although, as with the former (Sect. \ref{ionng}),
the monochromatic luminosity does not give the full picture.

\begin{figure*}
\centering 
\includegraphics[angle=270,scale=0.83]{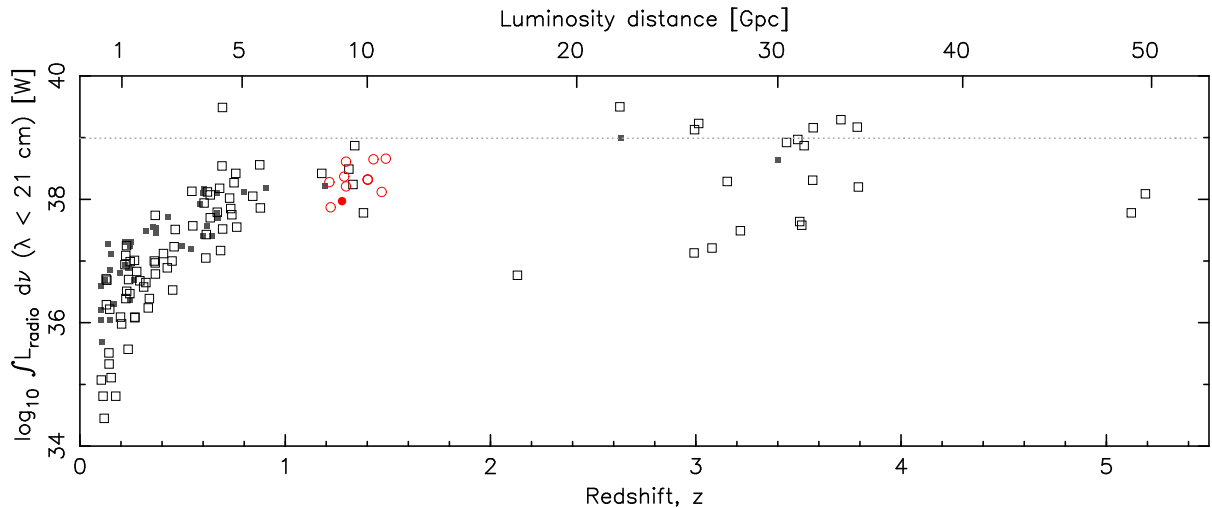}
\caption{The $\nu\geq1420$ MHz radio power versus redshift for the $z\geq0.1$ quasars and
radio galaxies searched in associated 21-cm absorption. The dotted horizontal line shows the highest value which there is a 21-cm detection.}
\label{pow-z}
\end{figure*}

Therefore, in Fig. \ref{pow-z} we show the radio power, $\int^{\nu_2}_{\nu_1}L_{\nu}\,d{\nu}$, where $\nu_1 = 1420$ MHz
and $\nu_2= 100$ GHz\footnote{Ideally, $\nu_2= \infty$ but we have restricted this to $10^{11}$ Hz, up to which we believe the
radio SED fits to be accurate \citep{cwsb12}. For the sources which do exhibit reasonable polynomial fits to the
SEDs, $\nu \gapp 10^{11}$~Hz photometries make very little contribution to the power.}, for all of the sources searched in 21-cm absorption.
From this, we see that while the radio power for J1545+475 is low compared to the rest of our targets, it is higher than some of the $z\gapp2$ sources.
In fact, most of these high redshift non-detections have radio powers below $9.8\times10^{38}$ W, the highest value for a 21-cm detection,
indicating that excitation to the
upper hyperfine level by $\lambda\leq 21$ cm photons cannot account for the non-detections at high redshift.  Given the 33\% detection
rate (40 out of 121) at $\leq9.8\times10^{38}$ W, the binomial probability of zero detections out of eight at $>9.8\times10^{38}$ W is 0.041,
which is significant at only $2.05\sigma$, assuming Gaussian statistics. By comparison, the significance
is $5.4\sigma$ for the  ionising photon rate \citep{cw12}.

\subsubsection{Redshifts}
\label{of}

Another possible reason for the non-detection of 21-cm absorption, in at least some of the targets, is inaccurate
optical redshifts, since in trawling the 24\,000 entries of the suitable radio catalogues (Sect. \ref{ss}), the optical
spectra were not checked.
As seen from Fig. \ref{spectra}, we achieve a range of $z_{\rm opt}\pm0.02$ which is large compared with any expected
statistical error in the optical redshifts. Of greater concern, is incorrectly classified spectral lines or noise in the
spectra misidentified as real emission lines resulting in spurious redshifts. A literature search revealed six of the
ten sources had published optical spectra and we checked the reliability of the redshifts of these objects. One source,
J1409--2657, had an incorrect published redshift of $z=2.43$ while the spectrum had high signal-to-noise ratio emission
lines clearly indicating a redshift of $z=1.43$. Two other objects, 0220--349 and 0357--264, were deemed to have redshifts
of marginal reliability, given the low signal-to-noise of the emission lines used in the classification. The remaining
three sources had a clear set of emission lines confirming the published redshift values. 
 
In order to account for all the possible factors leading to a non-detection, in Table \ref{redshifts} we flag the unreliable redshifts as incorrect ({\sf x})\footnote{Where the
spectrum is not published, we assume the derived redshift to be correct. For instance, \citet{vt95} quote $z = 1.277\pm0.002$ for J1545+4751.},
\begin{table} 
\centering
\begin{minipage}{85mm}
  \caption{The source references of the redshifts and the likelihood of a 21-cm detection, where `{\large \checkmark}'
    shows a favourable value, `{\sf x}' an unfavourable value and `--' an indeterminable value.  '$z_{\rm opt}$' is the
    optical redshift which receives a tick if deemed reliable and a cross if marginal (`--' signifies that the spectrum
    is not published and thus cannot be checked, see main text).  $\tau_{_{3\sigma}}$ is the optical depth limit, which
    receives a tick if the r.m.s. noise level obtained gives a limit of $N_{\rm HI} \ll1\times10^{19}\,(T_{\rm spin}/f)$
    \scm\ per channel (Table \ref{obs}), followed by the ionising photon rate, which receives a tick if
    $<2.6\times10^{56}$ s$^{-1}$, and the turnover frequency, which receives a tick if a turnover is evident in the
    radio SED (Table \ref{lums}).  The last column shows if a detection is deemed possible.  }
\begin{tabular}{@{}l c l  c  c c  c c @{}} 
\hline
\smallskip
Source                & Redshift & Ref.  & $z_{\rm opt}$ & $\tau_{_{3\sigma}}$ & Rate & $\nu_{_{\rm TO}}$ &  Det. \\
\hline
0220--349 &  1.49 & D97 & {\sf x} & {\sf x}  & -- &  {\large \checkmark} & N \\ 
J0301+0118  &  1.221& J02 & -- & {\large \checkmark} & {\large \checkmark}& {\sf x}  & N \\ 
0357--264 & 1.47& D97 & {\sf x}   & {\sf x}  & --  &  {\sf x}  &  N\\ 
0400--319 & 1.288& D97 &  {\large \checkmark} &  {\sf x}  & --  & {\large \checkmark}  & N \\
0511--220 &   1.296 & S89& {\large \checkmark} & {\large \checkmark} &{\large \checkmark}  & {\large \checkmark}  & Y\\
1004--018   & 1.2154& C04 & {\large \checkmark} & {\large \checkmark} & {\sf x} &  {\sf x}  & N\\
J1334--1150   &  1.402 & J02 & -- & {\large \checkmark} & -- &  {\sf x}  & N\\
J1409--2657  & 1.43& D97&  {\large \checkmark}& {\sf x}  & {\large \checkmark} & {\large \checkmark} & N\\
4C\,+04.51  &  1.996 & H91 &  --& {\large \checkmark} & {\large \checkmark}  & {\large \checkmark} & Y\\
J1545+4751  & 1.277& V95 & -- & {\large \checkmark} & {\large \checkmark} & {\large \checkmark} &  Y\\
\hline       
\end{tabular}
\label{redshifts}  
{References:  S89 -- \citet{sfk89}, H91 -- \citet{hobl91}, V95 -- \citet{vt95}, D97 -- \citet{dwf+97} [PHFS], J02 -- \citet{jws+02}, C04 -- \citet{csb+04}.}
\end{minipage}
\end{table} 
as well as flagging other conditions not favourable for a detection -- a poor sensitivity limit (Sect. \ref{sl}), an
ionising photon rate above 
the critical value (Sect. \ref{ionng}) and the absence of a detected turnover in
the SED, suggesting a low covering factor (Sect. \ref{cov}).  That is, a full row of ticks ({\large \checkmark}), from `$\tau_{_{3\sigma}}$' on, is
required for a possible detection (Y), whereas only one cross ({\sf x}) is required to rule out the possibility (N).
Summarising our targets thus, of the ten sources searched, we obtain only three in which the detection of 21-cm is
feasible.\footnote{Note that each of the sources for which the ionising photon rate is not known falls down by at least one {\sf x}, thus allowing us to rule out
  the possibility of a detection even if the ionising flux were below the critical value.} One of these is the detection, J1545+4751, which therefore
gives a $\approx30$\% detection rate. Given the small numbers, this is close to the 
rate expected at  $L_{\rm UV}\lapp10^{23}$ \WpHz\ \citep{cw10}. 

\section{Conclusions}

In the third in a series of surveys (following \citealt{cwwa11,cwsb12}) for redshifted \HI\ 21-cm in the hosts
of radio galaxies and quasars, we have searched ten, mostly flat spectrum, objects at redshifts of $z\sim1.2 - 1.5$. This corresponds to look-back
times of $\sim9$ Gyr, which is a range relatively unprobed by previous surveys for associated \HI\ 21-cm.
This has resulted in one detection, which at 
 $z= 1.2780$, after 4C\,+05.19 ($z= 2.6365$, \citealt{mcm98}) and B2\,0902+34  ($z=3.3980$, \citealt{ubc91}),
is the third highest redshift detection of associated 21-cm absorption. 

Of the remaining sources searched, one was 
completely lost to RFI, with another three being severely compromised (giving poor limits). 
After quantifying other pitfalls, which could prevent a detection, we estimate a detection rate of 30\%.
Although uncertain, this is consistent with the 40\% rate
generally found for those below the critical value (or luminosities of $L_{\rm UV}\lapp10^{23}$ \WpHz). 
This is in contrast to our previous survey of $z\gapp2$ sources, for which there were no detections. These sources did,
however, have much steeper spectra than the current sample, which may suggest that ineffective coverage of the
background emission could have contributed to the zero detection rate.

Despite our deliberate selection of optically faint sources, the one detection 
highlights the difficulties in searching for highly redshifted associated 21-cm absorption, where the selection 
of the necessarily faint objects introduces a bias towards objects for which the optical redshifts are poorly determined.
Furthermore, by probing these look-back times evolution in galactic morphologies may be an issue, where there is a larger fraction of smaller galaxies at high
 redshift \citep{bmce00,lf03}. For these the 
critical luminosity will be correspondingly lower \citep{cw12}, requiring the selection of even fainter optical sources
in order to detect associated 21-cm absorption. Thus, through their wide fields-of-view and bandwidths, making them
unreliant on optical redshifts, the next generation of radio telescopes (the Square Kilometre Array and its pathfinders)
will be ideal in finding the neutral gas currently missing in the hosts of high redshift  radio sources.

\section*{Acknowledgements}

We thank the staff of the GMRT who have made these observations possible. The GMRT is run by 
the National Centre for Radio Astrophysics of the Tata Institute of Fundamental Research
This research has made use of the NASA/IPAC Extragalactic Database
(NED) which is operated by the Jet Propulsion Laboratory, California
Institute of Technology, under contract with the National Aeronautics
and Space Administration. This research has also made use of NASA's
Astrophysics Data System Bibliographic Services.
The Centre for All-sky Astrophysics is an Australian Research Council Centre of Excellence, funded by grant CE110001020.



\begin{thebibliography}{}

\bibitem[\protect\citeauthoryear{{Allison}, {Curran}, {Emonts}, {Gereb},
  {Mahoney}, {Reeves}, {Sadler}, {Tanna} \& {Whiting}}{{Allison}
  et~al.}{2012}]{ace+12}
{Allison} J.~R.,  et al.,  2012,
  MNRAS, 423, 2601

\bibitem[\protect\citeauthoryear{{Baker}, {Mathlin}, {Churches} \&
  {Edmunds}}{{Baker} et~al.}{2000}]{bmce00}
{Baker} A.~C.,  {Mathlin} G.~P.,  {Churches} D.~K.,    {Edmunds} M.~G.,  2000,
  in Favata F.,  Kaas A.,   Wilson A.,  eds, Star Formation from the Small to
  the Large Scale, Vol.45 of ESA SP ESA Special Publication, {The Chemical
  Evolution of the Universe}.
Noordwijk, p.~21

\bibitem[\protect\citeauthoryear{{Catinella}, {Haynes}, {Giovanelli}, {Gardner}
  \& {Connolly}}{{Catinella} et~al.}{2008}]{chg+08}
{Catinella} B.,  {Haynes} M.~P.,  {Giovanelli} R.,  {Gardner} J.~P.,
  {Connolly} A.~J.,  2008, ApJ, 685, L13

\bibitem[\protect\citeauthoryear{{Croom}, {Smith}, {Boyle}, {Shanks}, {Miller},
  {Outram} \& {Loaring}}{{Croom} et~al.}{2004}]{csb+04}
{Croom} S.~M.,  {Smith} R.~J.,  {Boyle} B.~J.,  {Shanks} T.,  {Miller} L.,
  {Outram} P.~J.,    {Loaring} N.~S.,  2004, MNRAS, 349, 1397

\bibitem[\protect\citeauthoryear{Curran}{Curran}{2010}]{cur09a}
Curran S.~J.,  2010, MNRAS, 402, 2657

\bibitem[\protect\citeauthoryear{Curran}{Curran}{2012}]{cur12}
Curran S.~J.,  2012, ApJ, 748, L18

\bibitem[\protect\citeauthoryear{Curran, Tzanavaris, Murphy, Webb \&
  Pihlstr\"{o}m}{Curran et~al.}{2007}]{ctm+07}
Curran S.~J.,  Tzanavaris P.,  Murphy M.~T.,  Webb J.~K.,    Pihlstr\"{o}m
  Y.~M.,  2007, MNRAS, 381, L6

\bibitem[\protect\citeauthoryear{Curran, Webb, Murphy, Bandiera, Corbelli \&
  Flambaum}{Curran et~al.}{2002}]{cwbc01}
Curran S.~J.,  Webb J.~K.,  Murphy M.~T.,  Bandiera R.,  Corbelli E.,
  Flambaum V.~V.,  2002, PASA, 19, 455

\bibitem[\protect\citeauthoryear{Curran \& Whiting}{Curran \&
  Whiting}{2010}]{cw10}
Curran S.~J.,  Whiting M.~T.,  2010, ApJ, 712, 303

\bibitem[\protect\citeauthoryear{Curran \& Whiting}{Curran \&
  Whiting}{2012}]{cw12}
Curran S.~J.,  Whiting M.~T.,  2012, ApJ, 759, 117

\bibitem[\protect\citeauthoryear{Curran, Whiting, Murphy, Webb, Bignell,
  Polatidis, Wiklind, Francis \& Langston}{Curran et~al.}{2011a}]{cwm+10}
Curran S.~J.,  Whiting M.~T.,  Murphy M.~T.,  Webb J.~K.,  Bignell C.,
  Polatidis A.~G.,  Wiklind T.,  Francis P.,    Langston G.,  2011a, MNRAS, 413,
  1165

\bibitem[\protect\citeauthoryear{Curran, Whiting, Sadler \& Bignell}{Curran
  et~al.}{2012}]{cwsb12}
Curran S.~J.,  Whiting M.~T.,  Sadler E.~M.,  Bignell C.,  2012, MNRAS, in press (arXiv:1210.1886)

\bibitem[\protect\citeauthoryear{Curran, Whiting \& Webb}{Curran
  et~al.}{2009}]{cww09}
Curran S.~J.,  Whiting M.~T.,    Webb J.~K.,  2009, Proceedings of Science, 89,
  Chap. 11

\bibitem[\protect\citeauthoryear{Curran, Whiting, Webb \& Athreya}{Curran
  et~al.}{2011b}]{cwwa11}
Curran S.~J.,  Whiting M.~T.,  Webb J.~K.,    Athreya A.,  2011b, MNRAS, 414,
  L26

\bibitem[\protect\citeauthoryear{Curran, Whiting, Wiklind, Webb, Murphy \&
  Purcell}{Curran et~al.}{2008}]{cww+08}
Curran S.~J.,  Whiting M.~T.,  Wiklind T.,  Webb J.~K.,  Murphy M.~T.,
  Purcell C.~R.,  2008, MNRAS, 391, 765

\bibitem[\protect\citeauthoryear{{Drinkwater}, {Webster}, {Francis}, {Condon},
  {Ellison}, {Jauncey}, {Lovell}, {Peterson} \& {Savage}}{{Drinkwater}
  et~al.}{1997}]{dwf+97}
{Drinkwater} M.~J.,  {Webster} R.~L.,  {Francis} P.~J.,  {Condon} J.~J.,
  {Ellison} S.~L.,  {Jauncey} D.~L.,  {Lovell} J.,  {Peterson} B.~A.,
  {Savage} A.,  1997, MNRAS, 284, 85

\bibitem[\protect\citeauthoryear{{Fanti}, {Fanti}, {Schilizzi}, {Spencer}, {Nan
  Rendong}, {Parma}, {van Breugel} \& {Venturi}}{{Fanti} et~al.}{1990}]{ffs+90}
{Fanti} R.,  {Fanti} C.,  {Schilizzi} R.~T.,  {Spencer} R.~E.,  {Nan Rendong}
  {Parma} P.,  {van Breugel} W.~J.~M.,    {Venturi} T.,  1990, A\&A, 231, 333

\bibitem[\protect\citeauthoryear{{Grasha} \& {Darling}}{{Grasha} \&
  {Darling}}{2011}]{gd11}
{Grasha} K.,  {Darling} J.,  2011, in American Astronomical Society Meeting
  Abstracts Vol.~43, {A Search for Intrinsic HI 21-cm Absorption Toward Compact
  Radio Sources}.
p. 345.02

\bibitem[\protect\citeauthoryear{{Gupta}, {Salter}, {Saikia}, {Ghosh} \&
  {Jeyakumar}}{{Gupta} et~al.}{2006}]{gss+06}
{Gupta} N.,  {Salter} C.~J.,  {Saikia} D.~J.,  {Ghosh} T.,    {Jeyakumar} S.,
  2006, MNRAS, 373, 972

\bibitem[\protect\citeauthoryear{{Gupta}, {Srianand}, {Petitjean}, {Bergeron},
  {Noterdaeme} \& {Muzahid}}{{Gupta} et~al.}{2012}]{gsp+12}
{Gupta} N.,  {Srianand} R.,  {Petitjean} P.,  {Bergeron} J.,  {Noterdaeme} P.,
    {Muzahid} S.,  2012, A\&A, 544

\bibitem[\protect\citeauthoryear{{Heckman}, {O'Dea}, {Baum} \&
  {Laurikainen}}{{Heckman} et~al.}{1994}]{hobl91}
{Heckman} T.~M.,  {O'Dea} C.~P.,  {Baum} S.~A.,    {Laurikainen} E.,  1994,
  ApJ, 428, 65

\bibitem[\protect\citeauthoryear{{Helmboldt}, {Taylor}, {Tremblay},
  {Fassnacht}, {Walker}, {Myers}, {Sjouwerman}, {Pearson}, {Readhead},
  {Weintraub}, {Gehrels}, {Romani}, {Healey}, {Michelson}, {Blandford} \&
  {Cotter}}{{Helmboldt} et~al.}{2007}]{htt+07}
{Helmboldt} J.~F.,  et al.,  2007,
  ApJ, 658, 203

\bibitem[\protect\citeauthoryear{{Jackson}, {Wall}, {Shaver}, {Kellermann},
  {Hook} \& {Hawkins}}{{Jackson} et~al.}{2002}]{jws+02}
{Jackson} C.~A.,  {Wall} J.~V.,  {Shaver} P.~A.,  {Kellermann} K.~I.,  {Hook}
  I.~M.,    {Hawkins} M.~R.~S.,  2002, A\&A, 386, 97

\bibitem[\protect\citeauthoryear{{Lanfranchi} \& {Fria{\c c}a}}{{Lanfranchi} \&
  {Fria{\c c}a}}{2003}]{lf03}
{Lanfranchi} G.~A.,  {Fria{\c c}a} A.~C.~S.,  2003, MNRAS, 343, 481

\bibitem[\protect\citeauthoryear{{Machalski}}{{Machalski}}{1998}]{mac98}
{Machalski} J.,  1998, A\&AS, 128, 153

\bibitem[\protect\citeauthoryear{{Moore}, {Carilli} \& {Menten}}{{Moore}
  et~al.}{1999}]{mcm98}
{Moore} C.~B.,  {Carilli} C.~L.,    {Menten} K.~M.,  1999, ApJ, 510, L87

\bibitem[\protect\citeauthoryear{{Noterdaeme}, {Petitjean}, {Ledoux} \&
  {Srianand}}{{Noterdaeme} et~al.}{2009}]{npls09}
{Noterdaeme} P.,  {Petitjean} P.,  {Ledoux} C.,    {Srianand} R.,  2009, A\&A,
  505, 1087

\bibitem[\protect\citeauthoryear{Osterbrock}{Osterbrock}{1989}]{ost89}
Osterbrock D.~E.,  1989, Astrophysics of Gaseous Nebulae and Active Galactic
  Nuclei.
University Science Books, Mill Valley, California

\bibitem[\protect\citeauthoryear{Page, Symeonidis, Vieira, Altieri, Amblard \&
  Arumugam}{Page et~al.}{2012}]{psv+12}
Page M.~J., et al.,  2012, Nature, 485, 213

\bibitem[\protect\citeauthoryear{Prochaska, Hennawi \& Herbert-Fort}{Prochaska
  et~al.}{2008}]{phh07}
Prochaska J.~X.,  Hennawi J.~F.,    Herbert-Fort S.,  2008, ApJ, 675, 1002

\bibitem[\protect\citeauthoryear{{Purcell} \& {Field}}{{Purcell} \&
  {Field}}{1956}]{pf56}
{Purcell} E.~M.,  {Field} G.~B.,  1956, ApJ, 124, 542

\bibitem[\protect\citeauthoryear{Rao, Turnshek \& Nestor}{Rao
  et~al.}{2006}]{rtn05}
Rao S.,  Turnshek D.,    Nestor D.~B.,  2006, ApJ, 636, 610

\bibitem[\protect\citeauthoryear{{Stickel}, {Fried} \& {K\"{u}hr}}{{Stickel}
  et~al.}{1989}]{sfk89}
{Stickel} M.,  {Fried} J.~W.,    {K\"{u}hr} H.,  1989, A\&AS, 80, 103

\bibitem[\protect\citeauthoryear{{Stickel} \& {K\"{u}hr}}{{Stickel} \&
  {K\"{u}hr}}{1996}]{sk96b}
{Stickel} M.,  {K\"{u}hr} H.,  1996, A\&AS, 115, 11

\bibitem[\protect\citeauthoryear{{Taylor}, {Vermeulen}, {Pearson}, {Readhead},
  {Henstock}, {Browne} \& {Wilkinson}}{{Taylor} et~al.}{1994}]{tvp+94}
{Taylor} G.~B.,  {Vermeulen} R.~C.,  {Pearson} T.~J.,  {Readhead} A.~C.~S.,
  {Henstock} D.~R.,  {Browne} I.~W.~A.,    {Wilkinson} P.~N.,  1994, ApJS, 95,
  345

\bibitem[\protect\citeauthoryear{Uson, Bagri \& Cornwell}{Uson
  et~al.}{1991}]{ubc91}
Uson J.~M.,  Bagri D.~S.,    Cornwell T.~J.,  1991, PhRvL, 67, 3328

\bibitem[\protect\citeauthoryear{Vermeulen, Pihlstr\"{o}m, Tschager, de Vries,
  Conway, Barthel, Baum, Braun, Bremer, Miley, O'Dea, Roettgering, Schilizzi,
  Snellen \& Taylor}{Vermeulen et~al.}{2003}]{vpt+03}
Vermeulen R.~C., et al.,  2003, A\&A, 404, 861

\bibitem[\protect\citeauthoryear{{Vermeulen} \& {Taylor}}{{Vermeulen} \&
  {Taylor}}{1995}]{vt95}
{Vermeulen} R.~C.,  {Taylor} G.~B.,  1995, AJ, 109, 1983

\bibitem[\protect\citeauthoryear{{Wall} \& {Peacock}}{{Wall} \&
  {Peacock}}{1985}]{wp85}
{Wall} J.~V.,  {Peacock} J.~A.,  1985, MNRAS, 216, 173

\bibitem[\protect\citeauthoryear{{White}}{{White}}{1992}]{whi92a}
{White} G.~L.,  1992, PASA, 10, 140

\end{thebibliography}

\label{lastpage}

\end{document}